\documentclass[%twocolumn,
showpacs,preprintnumbers,amsmath,amssymb]{revtex4}

\begin{document}

\title{{\Large{\bf Aharonov-Bohm effect and geometric phases~\footnote{Invited talk given at Tonomura FIRST International Symposium on
gElectron Microscopy and Gauge Fieldsh, May 9-10, 2012, Tokyo, Japan.}}}\\ -- {\Large Exact and approximate topology--}}

\author{\large Kazuo Fujikawa}
\affiliation{\large
%Mathematical Physics Laboratory, 
RIKEN Nishina Center, Wako 351-0198, 
Japan}

\begin{abstract}
By analyzing an exactly solvable model in the second quantized formulation which allows a unified treatment of adiabatic and non-adiabatic geometric phases, it is shown that the topology of the adiabatic Berry's phase, which is characterized by the singularity associated with possible level crossing, is trivial in a precise sense. This topology of the geometric phase is quite different from the topology of the Aharonov-Bohm effect, where the topology is specified by the external local gauge field and it is exact for the slow as well as for the fast motion of the electron.

\end{abstract}

\maketitle

\large
\section{Aharonov-Bohm effect and geometric phases}
Akira Tonomura made important contributions to the studies of the Aharonov-Bohm effect and the double-slit experiment which is closely related to the analysis of geometric phases. Both of these effects are related to phases  and intereference in quantum mechanics. The phase in quantum mechanics is  also closely related to the notion of topology in mathematics.

The topology of the Aharonov-Bohm effect is provided by the external boundary condition for the gauge field~\cite{1959Aharonov},
 and the Aharonov-Bohm effect is best described by the path integral representation
\begin{eqnarray}
\langle \vec{x}_{f},T|\vec{x}_{i},0\rangle
&&=\int{\cal D}\vec{x}\exp\{\frac{i}{\hbar}\int_{0}^{T}[\frac{m\dot{\vec{x}}^{2}}{2}-e\vec{A}(\vec{x})\frac{d\vec{x}}{dt}]dt\}
\end{eqnarray}
for the propagation of an electron.

On the other hand, the geometric phase (or Berry's phase)~\cite{1975Higgins,
simon,1984Berry,1987Berry}
 for the electron placed in a rotating magnetic field $\vec{B}(t)$, which is solved exactly as shown below, is given by
\begin{eqnarray}\label{eq-exactamplitude1}
\psi_{\pm}(T)
&=&w_{\pm}(T)\exp\left[-\frac{i}{\hbar}\int_{0}^{T}dt
w_{\pm}^{\dagger}(t)\big(\hat{H}
-i\hbar\partial_{t}\big)w_{\pm}(t)\right]\nonumber\\
&=&w_{\pm}(T)\exp\left[-\frac{i}{\hbar}\int_{0}^{T}dt
w_{\pm}^{\dagger}(t)\hat{H}w_{\pm}(t)\right]\nonumber\\
&\times&\exp\left[-\frac{i}{\hbar}\int_{0}^{T}
\vec{{\cal A}}_{\pm}(\vec{B})\frac{d\vec{B}}{d t}dt\right],
\end{eqnarray}
where  
\begin{eqnarray}\label{connection}
\vec{{\cal A}}_{\pm}(\vec{B})\equiv w_{\pm}^{\dagger}(t)(-i\hbar\frac{\partial}{\partial \vec{B}})w_{\pm}(t)
\end{eqnarray}
gives an analogue of the gauge potential (or connection).
 
These two expressions are very similar, but the important difference is that the electron moves outside the magnetic field in the case of Aharonov-Bohm effect while the electron moves inside the magnetic field in the case of geometric phases. This difference suggests that the topology of these two phases, though similar, is fundamentally different. In fact, the topology of the Aharonov-Bohm effect is precise for the non-adiabatic  as well as adiabatic motion of the electron. On the other hand, it is shown that the topology of Berry's phase is valid only in the ideal adiabatic limit and it is lost once one moves away from ideal adiabaticity.  We would like to explain this difference in the following since it is not widely recognized.

\section{Second quantization and hidden gauge symmetry}
To analyze the topology of the geometric phase, one needs  a formulation which treats the adiabatic~\cite{1975Higgins,
simon,1984Berry,1987Berry}
 and nonadiabatic phases~\cite{1987Aharonov,
1988Samuel,1992Anandan}
 in a unified manner~\cite{2005Deguchi}. We start with the action
\begin{eqnarray}\label{eq-action1}
S&=&\int dtd^{3}x\left[\hat{\psi}^\dag(t.\vec{x})\Big(i\hbar\frac{\partial}{\partial
t}-\hat{H}(t)\Big)\hat{\psi}(t,\vec{x})\right]
\end{eqnarray}
for a time-dependent Hamiltonian $\hat{H}(t)$. We then expand
\begin{eqnarray}\label{psi}
\hat{\psi}(t,\vec{x})=\sum_{n}\hat{c}_{n}(t)v_{n}(t,\vec{x})
\end{eqnarray}
with $\int d^{3}xv_{n}^{\star}(t,\vec{x})v_{m}(t,\vec{x})=\delta_{n,m}$. 
For the fermion, we impose anti-commutator
$\big\{\hat{c}_l(t),\hat{c}_m^\dag(t)\big\}=\delta_{lm}$.
The Fock states are
defined by $|l\rangle=\hat{c}^\dag_l(0)|0\rangle$.

By inserting the expansion into the action $S$, we have
\begin{eqnarray}
S&=&\int dt \{\sum_{n}\hat{c}^{\dagger}_{n}(t)i\hbar\partial_{t}\hat{c}_{n}(t)\nonumber\\
&-&\sum_{n,m}\int d^{3}x[ v^{\star}_{n}(t,\vec{x})\hat{H}(t)v_{m}(t,\vec{x})
-v^{\star}_{n}(t,\vec{x})i\hbar\partial_{t}v_{m}(t,\vec{x})]\hat{c}^{\dagger}_{n}(t)\hat{c}_{m}(t)\},
\end{eqnarray}
and the appearance of "geometric phase" in the last term is automatic.

The solution of the conventional Schr\"{o}dinger equation
with the initial condition $\psi(0,\vec{x})=v_{n}(0,\vec{x})$
is given by $\psi_{n}(t,\vec{x})=\langle 0|\hat{\psi}(t,\vec{x})\hat{c}^{\dagger}_{n}(0)|0\rangle$.
This second quantized formulation contains the following {\em gauge} (or redundant) freedom~\cite{2005Fujikawa}
\begin{eqnarray}\label{hidden-gauge}
\hat{c}_{n}(t)\rightarrow e^{-i\alpha_{n}(t)}\hat{c}_{n}(t),
\hspace{5mm} v_{n}(t)\rightarrow e^{i\alpha_{n}(t)}v_{n}(t)
\end{eqnarray}
which keeps $\hat{\psi}(t,\vec{x})$ in (\ref{psi}) invariant, with the phase freedom $\{\alpha_{n}(t)\}$ being arbitrary functions of time. 
Under this {\em hidden gauge transformation}, the Schr\"{o}dinger amplitude is transformed as
\begin{eqnarray}
\psi_{n}(t,\vec{x})=\langle 0|\hat{\psi}(t,\vec{x})\hat{c}^{\dagger}_{n}(0)|0\rangle\rightarrow e^{i\alpha_{n}(0)}\psi_{n}(t,\vec{x}),
\end{eqnarray}
namely, the ray representation of the state vector is induced.  
One may ask what is the physical implication of this hidden gauge symmetry?
The answer is "it controls all the geometric phases, either adiabatic or non-adiabatic"~\cite{2007Fujikawa}.
In the analysis of geometric phases, it is crucial to note that the combination 
$\psi_{n}^{\star}(0,\vec{x})\psi_{n}(t,\vec{x})$
is manifestly gauge invariant.

\section{Exactly solvable example and geometric phase}

We consider the motion of a spin inside the rotating magnetic field 
\begin{eqnarray}
 {\bf B}(t)=B\big(\sin\theta \cos\varphi(t),
\sin\theta\sin\varphi(t),\cos\theta\big)
\end{eqnarray}
 and $\varphi(t) =\omega_0t$ with a constant $\omega_0$.
The action  is written as 
\begin{eqnarray}\label{eq-action1}
S&=&\int dt\left[\hat{\psi}^\dag(t)\Big(i\hbar\frac{\partial}{\partial
t}+{\bf B}\cdot {\bf \sigma}/2\Big)\hat{\psi}(t)\right],
\end{eqnarray}
with ${\bf \sigma}$  Pauli matrices.
The field operator is expanded as $\hat{\psi}(t,\vec{x})=\sum_{l=\pm}\hat{c}_l(t)w_l(t)$
 with the anti-commutation relation,
$\big\{\hat{c}_l(t),\hat{c}_m^\dag(t)\big\}=\delta_{lm}$.

The effective Hamiltonian for the above spin system is exactly diagonalized if one defines 
\begin{eqnarray}
w_{+}(t)=\left(\begin{array}{c}
             e^{-i\varphi(t)}\cos\frac{\vartheta}{2}\\
            \sin\frac{\vartheta}{2}
            \end{array}\right),\
w_{-}(t)=\left(\begin{array}{c}
             e^{-i\varphi(t)}\sin\frac{\vartheta}{2}\\
            -\cos\frac{\vartheta}{2}
            \end{array}\right)
\end{eqnarray}
with $\vartheta=\theta-\theta_0$, and the constant parameter
$\theta_0$ is defined by
\begin{equation}\label{eq-xi}
\tan\theta_0=\frac{\hbar\omega_0 \sin\theta}{B+\hbar\omega_0\cos\theta}.
\end{equation}
The effective Hamiltonian is then written as $\hat{\mathcal{H}}_{\text{eff}}(t)\equiv\sum_lE_l\hat{c}_l^\dag(0)\hat{c}_l(0)$
with time-independent effective energy eigenvalues
\begin{eqnarray}\label{eq-Epm}
E_\pm&=&w_{\pm}^{\dagger}(t')\big(\hat{H}
-i\hbar\partial_{t'}\big)w_{\pm}(t')\nonumber\\
&=&
\mp\frac{1}{2}B\cos\theta_0-\frac{1}{2}\hbar\omega_0\big[1\pm\cos(\theta-\theta_0)\big].
\end{eqnarray}

The {\em exact} solution of the Schr\"{o}dinger equation is given by~\cite{2007Fujikawa}
\begin{eqnarray}\label{eq-exactamplitude}
\psi_{\pm}(t)&=&\langle
0|\hat{\psi}(t)\hat{c}^\dag_\pm(0)|0\rangle\nonumber\\
&=&w_{\pm}(t)\exp\left[-\frac{i}{\hbar}\int_{0}^{t}dt'
w_{\pm}^{\dagger}(t')\big(\hat{H}
-i\hbar\partial_{t'}\big)w_{\pm}(t')\right]\nonumber\\
&=&w_{\pm}(t)\exp\left[-\frac{i}{\hbar}E_{\pm}t\right],
\end{eqnarray}
where the exponent has been calculated in Eq.(\ref{eq-Epm}). 

The basis vectors satisfy $w_{\pm}(T)=w_{\pm}(0)$ with the period $T=2\pi/\omega_0$. The
solution is thus {\em cyclic} (namely, periodic up to a phase freedom) and, as an exact solution, it is applicable to the
non-adiabatic case also.

%Incidentally, for an arbitrary time-dependent ${\bf B}(t)$, {\em any exact} solution of the Schr\"{o}dinger equation can be written in the form of Eq.(\ref{eq-exactamplitude1}), if one chooses basis vectors $w_{\pm}(t)$ suitably. 

\subsection{ Adiabatic limit} 
The adiabatic limit is defined by 
$|\hbar\omega_0/ B|\ll 1$ for which the parameter $\theta_0 \rightarrow 0$ in Eq.(\ref{eq-xi}), and the exact Schr\"{o}dinger amplitude (\ref{eq-exactamplitude}) approaches
\begin{eqnarray}\label{eq-exactamplitude2}
\psi_{\pm}(t)
%&=&w_{\pm}(t)\exp\left[-\frac{i}{\hbar}[-\frac{1}{2}\hbar\omega_0\big(1\pm\cos(\theta-\theta_0)\big)t\right]\nonumber\\
%&&\hspace{6mm}\times\exp\left[-\frac{i}{\hbar}[
%\mp\frac{1}{2}B\cos\theta_0]t\right]\\
&\Rightarrow&w_{\pm}(t)\exp\left[\frac{i}{2}[\omega_0\big(1\pm\cos\theta \big)t\right] \exp\left[-\frac{i}{\hbar}[
\mp\frac{1}{2}B]t\right]
\end{eqnarray}
where the first phase factor is called  {\em geometric phase} and
 the second phase factor as {\em dynamical phase}.

The conventional geometric phase or "Berry's phase"  
\begin{eqnarray}
\exp{[i\pi(1\pm\cos\theta)]}
\end{eqnarray}
is recovered after one cycle $t=T=2\pi/\omega_{0}$ of the motion. 
This Berry's phase is known to have a {\em topological meaning} as the phase generated by a magnetic monopole located at the origin of the parameter space ${\bf B}$~\cite{1987Berry}.

Note that the dynamical phase in (\ref{eq-exactamplitude2}) vanishes at ${\bf B}=0$, namely, the {\em level crossing} appears in the conventional adiabatic approximation. 

We note that, in the generic case (\ref{eq-exactamplitude}) with period $T=\frac{2\pi}{\omega_{0}}$, one can in principle
 measure $\psi^{\dagger}_{+}(0)\psi_{+}(T)$ by looking at the 
interference~\cite{2007Fujikawa} 
\begin{eqnarray}
|\psi_{+}(T)+\psi_{+}(0)|^{2}
&=&2|\psi_{+}(0)|^{2}+2{\rm Re}\psi^{\dagger}_{+}(0)\psi_{+}(T)
\nonumber\\
&=&2+2\cos[(\mu B
\cos\theta_0)T
-\frac{1}{2}\Omega_{+}],
\end{eqnarray}
where the geometric phase
\begin{eqnarray}
\Omega_{+}=2\pi [1-\cos(\theta-\theta_0)]
\end{eqnarray}
stands for the solid angle drawn by 
$w_{+}^{\dagger}(t)\vec{\sigma}w_{+}(t)$.

\subsection{ Non-adiabatic limit} 
The non-adiabatic limit is defined by $|\hbar\omega_0/ B|\gg 1$, and thus
$\theta_0 \rightarrow \theta$ in Eq.(\ref{eq-xi}) so that the geometric phase vanishes in the exact amplitude (\ref{eq-exactamplitude}), 
\begin{eqnarray}
\exp\left[-\frac{i}{\hbar}[-\frac{1}{2}\hbar\omega_0\big(1\pm\cos(\theta-\theta_0)\big)\frac{2\pi}{\omega_{0}}\right]=1.
\end{eqnarray}
The formal gauge connection in (\ref{connection}) also vanishes.
Namely, the {\em adiabatic Berry's phase is smoothly connected to the trivial phase inside the exact solution} and thus the topology of Berry's phase is trivial in a precise sense. 
In our unified formulation of adiabatic and non-adiabatic phases, we can analyze a transitional
region from the adiabatic limit to the non-adiabatic region in a reliable way, which was not possible in the past formulation.

\section{Conclusion}

The present second quantized approach allows a unified treatment of all the geometric phases, either adiabatic or non-adiabatic, and thus one can analyze the 
transitional region from adiabatic to non-adiabatic phases in a reliable way.
One then recognizes that the topology of the adiabatic Berry's phase is actually trivial, in contrast to the topology of Aharanov-Bohm effect which is exact for the fast as well as slow motion of the electron.

The analyses of other aspects of geometric phases form a point of view of second quantization are found in the review~\cite{2009Fujikawa} with further references.

\end{document}